# Exploring Engagement and Perceived Learning Outcomes in an Immersive Flipped Learning Context

Mehrasa Alizadeh
m-alizadeh@otemon.ac.jp
Otemon Gakuin University, Japan

*Abstract:* The flipped classroom model has been widely acknowledged as a practical pedagogical approach to enhancing student engagement and learning. However, it faces challenges such as improving student interaction with learning content and peers, particularly in Japanese universities where digital technologies are not always fully utilized. To address these challenges and identify potential solutions, a case study was conducted in which an online flipped course on academic skills was developed and implemented in an immersive virtual environment. The primary objective during this initial phase was not to establish a causal relationship between the use of immersive flipped learning and students' engagement and perceived learning outcomes. Instead, this initiative aimed to explore the benefits and challenges of the immersive flipped learning approach in relation to students' online engagement and their perceived learning outcomes. Following a mixed-methods research approach, quantitative and qualitative data were collected through a survey (N=50) and students' reflective reports (N=80). The study revealed high levels of student engagement and perceived learning outcomes, although it also identified areas needing improvement, particularly in supporting student interactions in the target language. Despite the exploratory nature of this study, the findings suggest that a well-designed flipped learning approach, set in an engaging immersive environment, can significantly enhance student engagement, thereby supporting the learning process. When creating an immersive flipped learning course, educators should incorporate best practices from the literature on both flipped learning and immersive learning design to ensure optimal learning outcomes. The findings of this study can serve as a valuable resource for educators seeking to design engaging and effective remote learning experiences. Further research is needed to explore the scalability and sustainability of the immersive flipped learning approach and its impact on different student populations and in other delivery formats.

*Keywords*: Flipped learning, Immersive learning, Engagement, Perceived learning outcomes,

# INTRODUCTION

**STUDENT ENGAGEMENT**

Providing quality education at the university level is essential for shaping the future of individuals and society. To achieve this goal, universities must ensure that the learning experience is not just informative but also engaging, inspiring students to be active learners who explore ideas and challenge assumptions. Universities can instill a sense of ownership and agency in students by creating an interactive learning environment that promotes collaboration and discussion, encourages critical thinking and problem-solving skills, and facilitates student-centered activities. This, in turn, can lead to better learning outcomes and more successful graduates. Therefore, promoting student engagement is a key factor in providing quality education at the university level.

Student engagement refers to the degree to which learners are involved in activities and conditions that lead to high-quality learning (Leach, 2014). Engaged students are more likely to attend classes regularly, complete assignments on time, interact with their peers and seek help from instructors when needed. They are also more likely to retain information, apply knowledge in real-world situations, and achieve better academic outcomes. Student engagement has been widely theorized and researched in the field of education (Kahu, 2013). Researchers have developed various models and frameworks to define and measure engagement, such as Fredricks et al.'s (2004) model of behavioral, emotional, and cognitive engagement, Reeve's concept of agentic engagement (Reeve, 2013), and more recently, the online engagement model by Redmond et al. (2018), to name a few. Studies have shown a positive relationship between student engagement and academic achievement (Carini et al., 2006; Guo et al., 2022). When students are engaged, they are more likely to perform well on exams, earn higher grades, and persist through university. Conversely, disengaged students are more likely to drop out, experience academic difficulties, and struggle to find employment after graduation. Universities worldwide have recognized student engagement's importance and implemented various strategies to promote it. These strategies include active learning techniques and student-centered pedagogy (Exeter et al., 2010; Kember, 2009). Moreover, Information and Communications Technology (ICT) has enabled institutions to create new opportunities for engagement, such as online and blended learning (Bond et al., 2020; Cole et al., 2019; Sun & Rueda, 2012; Yousaf et al., 2022).

**FLIPPED CLASSROOM MODEL**

One alternative approach that has gained popularity in recent years is the flipped classroom model. The flipped classroom model is a pedagogical approach that has proven beneficial in enhancing student engagement and learning (O'Flaherty & Phillips, 2015). It is an approach where students are assigned to watch pre-recorded lectures or video presentations at home prior to attending a face-to-face class. This frees up in-class time and allows students to focus on applying the concepts learned from the online materials through interactive group discussions, problem-solving exercises, and hands-on activities. The approach has been shown to improve student engagement and learning, as it allows them to take control of their own learning pace and allows them to co-construct and apply knowledge as they work collaboratively with peers and the teacher during class time (Bui et al., 2022; Divjak et al., 2022; Tang et al., 2020; Zainuddin et al., 2021). Additionally, the flipped classroom model can lead to better retention of information, as students are given multiple opportunities to engage with the material and apply it in different contexts. As such, the flipped classroom model is gaining popularity in educational institutions worldwide to engage students and optimize their learning outcomes.

While the flipped classroom model effectively enhances student engagement and learning, it has its shortcomings. One major challenge is that not all students may be motivated to engage with pre-class materials or prepare adequately for class, which could undermine the effectiveness of in-class activities (Fischer & Yang, 2022). Another risk is that students who cannot keep up with pre-class materials may fall behind and feel overwhelmed, leading to decreased engagement and motivation (Lee & Choi, 2018). Additionally, the flipped classroom model requires significant planning and preparation from instructors, which can be time-consuming and challenging, especially in the beginning stages (Giannakos et al., 2014). Lastly, despite the frequently reported benefits of flipped classrooms, reviews of the literature have identified contradictory results and challenges that need to be further investigated (Akçayır & Akçayır, 2018; Linling & Abdullah, 2023; Lo, 2023; Turan & Akdag-Cimen, 2019; Zainuddin & Halili, 2016). Akçayır and Akçayır (2018) identified five categories of challenges associated with the flipped classroom, including pedagogical, students' and teachers' perspectives, technical and technological, and other challenges. Their study revealed that most of these challenges stem from inadequate student preparation before class and insufficient guidance for self-regulated learning. More recently, Lo (2023) has highlighted the difficulties associated with online flipped learning during class, such as inadequate student participation, instructors' inability to monitor students via the screen, and a lack of practical, hands-on learning opportunities. These latest findings underscore the need for further research into effective strategies for designing hands-on flipped learning experiences, leveraging new technologies to facilitate student response and collaboration, and integrating formative assessment and feedback into the learning design.

While these challenges can certainly be significant, it is worth noting that various factors, including cultural norms and educational practices, may influence them. In the context of Japan, for example, there are some unique obstacles to implementing the flipped classroom model. Design frameworks of the flipped classroom argue that educational technology facilitates the planning and implementation of flipped classrooms. It can empower learners by giving them more opportunities for interaction and collaboration (Kim et al., 2014; Lo, 2023). Despite technology's major role in supporting student interaction and engagement and improving learning outcomes in flipped classrooms, progress in ICT integration at Japanese universities has generally been relatively slow (Mehran et al., 2017). Prior to the spread of COVID-19 in Japan, only a small number of universities had endorsed ICT integration by establishing infrastructure for technology-enhanced learning, including Learning Management Systems (LMS), Open Educational Resources (OER), Massive Open Online Courses (MOOCs), and e-portfolios. Many universities, by and large, have been slow to adopt and utilize digital technologies in their academic programs, as noted by Funamori (2017). The COVID-19 pandemic highlighted the lack of preparedness among most Japanese universities for online education. With emergency remote teaching thrust upon them, many instructors resorted to creating video recordings of their lectures and providing them to students on demand. While subsequent studies have shown that Japanese students appreciate the flexibility of this approach to online learning (Toda, 2021), limited interaction between students and teachers and among students remains a significant challenge. It is worth noting that this limitation is more pronounced in Japan due to cultural and historical factors that make asynchronous communication on discussion forums less popular among Japanese students (Nielsen, 2013; Yu & Hu, 2022). Maintaining student engagement and satisfaction has been a significant concern in Japan's online higher education context. Consequently, despite the benefits of online/blended learning and the flipped classroom model, most higher education programs returned to in-person instruction soon after the pandemic, highlighting the limitations of existing approaches for maintaining student engagement and satisfaction.

To address these obstacles and capitalize on the advantages of the flipped classroom model in online and blended learning environments, this study aims to design, implement, and evaluate an immersive flipped course on academic skills for university students. The research represents a novel effort to merge three pedagogies—flipped learning, collaborative learning, and immersive learning—to improve students' online participation and perceived learning outcomes. The paper presents a case study in which an online flipped course was developed incorporating collaborative tasks and implemented in an immersive virtual setting.

This research aims to examine the potential relationship of the immersive flipped learning approach with student online engagement and perceived learning outcomes. The following two research questions guided the case study:
1. To what extent are students engaged in this immersive flipped online course?
2. What learning outcomes do students perceive they have achieved in this immersive flipped online course?

The study provides valuable insights into the potential benefits of the immersive flipped learning approach for online education by investigating the two research questions above. It presents a case study and implications for educators seeking to design immersive flipped learning experiences. In the following sections, we will delve into the course's design and components, the online platform used to deliver and run the course, and the students' perspectives of the immersive flipped learning experience.

## RESEARCH DESIGN & METHODS

**COURSE COMPONENTS AND LEARNING ACTIVITIES**

The course titled "Integrated English (Academic Skills)" is an elective course offered to all undergraduate students at a national university in western Japan. It is part of the university's general education curriculum that aims to develop students' fundamental skills necessary to engage in academic activities in their second language (L2). Multiple instructors teach the course, each offering their unique approach to covering the general course topics of (a) note taking, (b) skimming and scanning, (c) critical reading and thinking, (d) writing academic essays, and (e) citing sources and compiling a bibliography. To develop these overarching skills, the author structured her syllabus into four parts: (1) academic reading, (2) academic writing, (3) academic presentation, and (4) additional skills. The fifteen-week course covers various topics, curated after a thorough analysis of syllabi from esteemed international universities and consultation with literature on academic and study skills for English-as-a-second/foreign-language students. Table 1 summarizes the course components, learning contents, and activities and their use and significance.

As Table 1 shows, the course was hosted on Gather (https://www.gather.town/), a 2D Metaverse platform (Wu et al., 2024) that offers unique features such as avatars, spatial audio and video interactions, content embedding, and customizable environments. The platform also allows for easy navigation and movement across breakout zones, with all participants being visible in each zone but only able to communicate with those in the same private zone. Figure 1 displays a screenshot of various areas within the larger Gather space shown on the left, along with a close-up view of several students in a breakout zone on the right. The course was structured around group activities to foster the students' collaborative skills. As the course was delivered entirely online, apps like Google Apps and Padlet were extensively utilized to encourage collaboration among small groups of students.

**PARTICIPANTS AND DATA COLLECTION**

In this case study, the participants consisted of three groups of students enrolled in a university course titled Integrated English (Academic Skills) taught by the author. The total enrollment for these three groups comprised 87 first- and second-year undergraduate students from diverse backgrounds and fields, including humanities, social sciences, and science and engineering. Participation was voluntary and anonymous. Following a mixed-methods research approach (Riazi & Candlin, 2014), both quantitative and qualitative data were collected from the students to evaluate the course. For quantitative evaluation, a five-point Likert scale survey was distributed online to all students at the end of the semester. The questionnaire was originally developed and validated by Hoi (2022) and measures students' perception of an engaging online learning environment across seven dimensions: course clarity, student connectedness, course structure, provision of choice, teaching relevance, teacher emotional support, and teacher presence. The author translated it into Japanese and independently revised it by two native Japanese speakers, one with expertise in language education and the other in educational technology. The author reviewed and incorporated their feedback to create a final electronic version of the questionnaire using Microsoft Forms. It was distributed to all students enrolled in three classes (see Appendix).

Table 1

*Course Components, Use, and Significance*

| Course Components | Use and Significance |
|---|---|
| Course LMS | The LMS served as the central hub for the course, where all the necessary materials, links, and announcements were shared. To ensure that students kept up with the coursework, they were given weekly online assignments that had to be submitted through the LMS. Additionally, the LMS provided an organized platform for students to access all the group work content discussed during weekly lessons, and it played a vital role in facilitating effective communication and content sharing. |
| Gather Virtual Platform | Gather, a 2D Metaverse platform was used to conduct synchronous lessons in a gamified environment. This provided a safe and enjoyable space for students to collaborate on joint tasks and achieve shared goals. Additionally, the platform was utilized to share pre-class materials with students to optimize interactive learning experiences during virtual meetings. This innovative platform helped facilitate an engaging and productive learning environment. |
| Padlet | Padlet was used as a platform for posting questions related to pre-class content, and students could answer them in real-time using the voting function and/or the comment box. Although communication was limited to writing, this method proved to be an effective way of reviewing lesson contents prior to group work. |
| Open Educational Resources | The course materials primarily consisted of open educational resources sourced from freely available websites and YouTube. Before every lesson, students were provided with links to these materials on both the Gather platform and the course LMS, enabling them to preview and later review the contents at their convenience. |
| Google Apps | The collaborative learning and problem-solving process relied heavily on Google apps, particularly Google Docs and Slides. These tools were frequently used to present group tasks to students and solicit their ideas and solutions. |

As part of the qualitative evaluation of the course, the author analyzed students' reflective reports, which accounted for 40% of their final grades. The students were asked to write a report of 500 to 800 words in which they would explain how the course helped (or alternatively failed to help) them improve their (a) academic skills, (b) collaboration skills, and (c) digital literacy, with supporting details. The submitted reports were analyzed qualitatively using MAXQDA version 2022. Based on the report instructions, the author employed a coding scheme of three overarching themes. These themes were used as the main categories for analysis and included the impact of the course in improving (1) academic skills, (2) collaboration skills, and (3) digital literacy.

*Figure 1*

*Screenshot of Students in Gather 2D Metaverse*

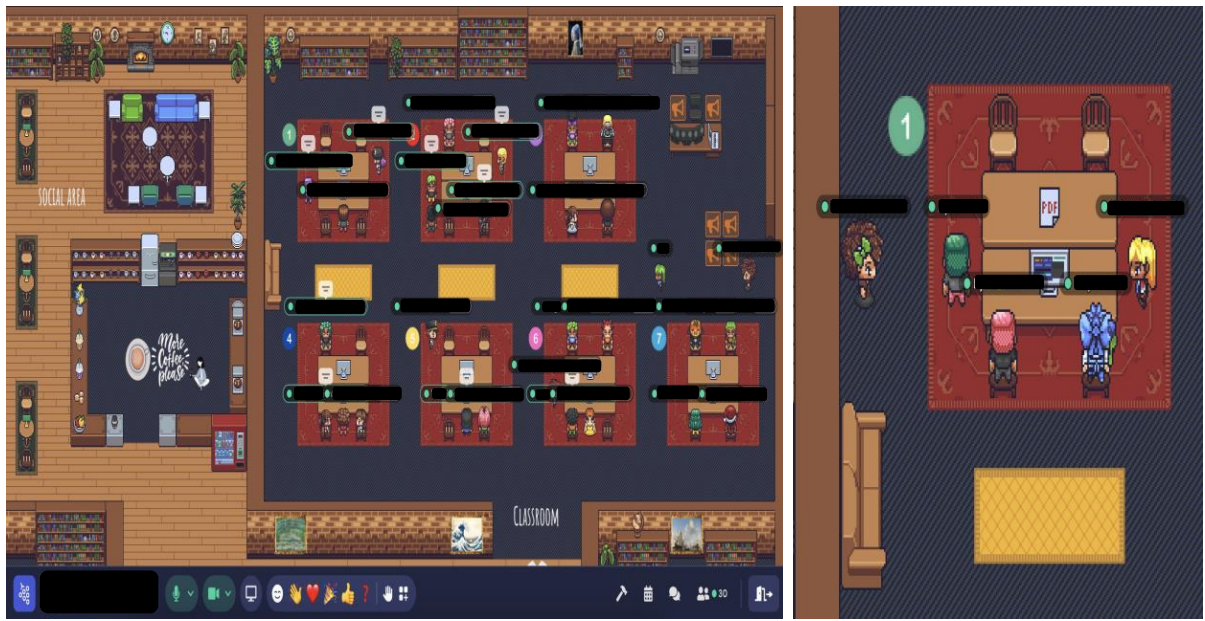

# COURSE EVALUATION RESULTS

The immersive flipped course evaluation followed a mixed-methods approach. This involved collecting quantitative survey data from students to measure their online engagement and qualitatively analyzing the content of their final reports to examine perceived learning outcomes.

**STUDENT ENGAGEMENT: SURVEY**

Out of the initial pool of 87 students, 50 voluntarily participated in the survey with complete anonymity and privacy protection measures in place. The descriptive statistics in Table 2 show the mean, median, and standard deviation values based on the group ratings. A reliability check was conducted using IBM SPSS version 26 to verify the internal consistency of the student's responses. The Cronbach's Alpha test produced an index of .89, signifying a high level of reliability.

Eyeballing the values in Table 2 indicates that the students held a highly favorable view of the course and reported a strong engagement in the online environment. A one-sample Wilcoxon signed-rank test was conducted to investigate this observation further using IBM SPSS version 26 for Windows.

The test compared the median values of the survey responses to a hypothetical median of 3, in this case, the scale midpoint, to statistically test the hypothesis that most responses would be "agree" or "strongly agree" (i.e., greater than "neutral"). All the 25 null hypotheses—The median of item x equals 3.00—were rejected at $p < .05$, confirming that most respondents perceived the online course as satisfactory and highly engaging.

Table 2

*Descriptive Statistics of the 5-Point Likert Scale Survey (N=50)*

| Item Dimension | Item* | Mean | Median | Std. Deviation | Sig. |
|---|---|---|---|---|---|
| Course clarity | Item1 | 4.80 | 5.00 | 0.40 | .000*** |
|  | Item2 | 4.56 | 5.00 | 0.58 | .000*** |
|  | Item3 | 4.70 | 5.00 | 0.51 | .000*** |
|  | Item4 | 4.74 | 5.00 | 0.49 | .000*** |
|  | Item5 | 4.68 | 5.00 | 0.51 | .000*** |
| Student connectedness | Item6 | 4.52 | 5.00 | 0.58 | .000*** |
|  | Item7 | 4.14 | 4.00 | 0.83 | .000*** |
|  | Item8 | 3.62 | 4.00 | 0.95 | .000*** |
|  | Item9 | 4.08 | 4.00 | 0.83 | .000*** |
| Course structure | Item10 | 4.50 | 5.00 | 0.65 | .000*** |
|  | Item11 | 4.60 | 5.00 | 0.53 | .000*** |
|  | Item12 | 4.32 | 4.00 | 0.71 | .000*** |
| Provision of choice | Item13 | 4.40 | 5.00 | 0.90 | .000*** |
|  | Item14 | 3.84 | 4.00 | 0.62 | .000*** |
|  | Item15 | 4.14 | 4.00 | 0.67 | .000*** |
|  | Item16 | 3.56 | 4.00 | 1.05 | .001*** |
| Teaching relevance | Item17 | 3.92 | 4.00 | 0.78 | .000*** |
|  | Item18 | 3.66 | 4.00 | 0.96 | .000*** |
|  | Item19 | 4.00 | 4.00 | 0.95 | .000*** |
| Teacher emotional support | Item20 | 4.74 | 5.00 | 0.49 | .000*** |
|  | Item21 | 4.80 | 5.00 | 0.45 | .000*** |
|  | Item22 | 4.78 | 5.00 | 0.46 | .000*** |
| Teacher presence | Item23 | 4.88 | 5.00 | 0.33 | .000*** |
|  | Item24 | 4.86 | 5.00 | 0.35 | .000*** |
|  | Item25 | 4.84 | 5.00 | 0.42 | .019*** |

*Note*. Check the Appendix for items 1 to 25.

**PERCEIVED LEARNING OUTCOMES: STUDENT REPORTS**

As part of the qualitative evaluation of the course, the author analyzed students' reflective reports. They were asked to write a report in which they explained how the course helped them or not in improving their (a) academic skills, (b) collaboration skills, and (c) digital literacy, with supporting details. Out of 87 students, 80 submitted the final report, and their reports were analyzed qualitatively using MAXQDA version 2022, a proprietary software program for qualitative data analysis.

The students unanimously endorsed the course for its effectiveness in improving their academic reading, writing, and presentation skills and providing hands-on experience through group activities. Before the course, many students were unfamiliar with reading techniques such as skimming and scanning, resulting in time management issues during English proficiency tests like the TOEFL.

However, after learning these strategies, several students reported an improvement in their test scores and even applied them to reading academic texts in Japanese. As one student wrote, "I was able to make full use of this technique in the TOEFL I took recently, and I was able to apply it not only to English sentences but also to Japanese sentences."

As academic writing was a significant component of the course, all students reflected on it in their reports and expressed gratitude for the opportunity to learn how to write academic reports in English. Many students had prior experience writing reports in their first language but were unfamiliar with the structure and organization of reports in English. In addition, several students emphasized the significance of understanding citation, referencing, and plagiarism. One student succinctly summarized the importance of learning how to write academic reports:

> "I felt that three things were significant in academic reports. They are understanding the importance of the introduction and conclusion, the importance of the topic sentence in each paragraph, and the importance of absolutely avoiding plagiarism and writing academic reports with academic integrity."

Similarly, the students acknowledged the significance of gaining knowledge about diverse forms of academic presentation, including unconventional formats like the Pecha Kucha. They expressed their eagerness to attempt giving a Pecha Kucha or poster presentation in the future. Regarding all three academic skills under scrutiny, the students valued the fact that these skills were transferable and could be utilized in other courses, study abroad programs, and future research and careers.

Many students lacked experience in group work in terms of collaboration skills. This was partially due to the COVID-19 pandemic, which disrupted school education and prevented teachers from including group tasks that involved close contact. Initially, it was challenging for most students to break the ice and work towards a shared goal with total strangers remotely. However, after several practice sessions in the gamified environment of the Metaverse, most improved their ability to initiate conversations and express their opinions. Some even developed leadership skills and guided group discussions.

Fortunately, the Metaverse offered these collaborative opportunities within a safe environment. Nonetheless, most conversations took place in Japanese, which some students regretted. Although the author encouraged students to communicate in English, they preferred Japanese as it was their shared language and a more accessible option. Several students wished they could have used more English in class. One student suggested limiting discussions to English as much as possible, while another expressed regret that they could not use English during group work.

Since the course was fully online and incorporated various digital tools, the students reflected on how it potentially helped them develop their digital literacy. The students not only mentioned the online environment of the course and digital platforms such as Gather, Google apps, and Padlet, but they also highlighted additional skills such as database searching and writing professional emails in English as part of their digital literacy skill set. Some students even acknowledged the importance of digital note-taking, which was only briefly mentioned in class. The transferability of these skills for future research and career opportunities was also frequently mentioned by numerous students.

## DISCUSSION

This study investigated the engagement and perceived learning outcomes of first and second-year undergraduate students in an immersive flipped course hosted remotely on a 2D Metaverse platform. The results of this study suggest that the course was effective in engaging students and helping them improve their academic skills in reading, writing, presentation, collaboration, and digital literacy. The students' positive evaluation of the course was reflected in the survey responses, which showed that they perceived the course as highly engaging and satisfactory.

The students also reported significant academic improvement, particularly in reading, writing, and presentation. The fact that these skills were transferable and applicable to their future research and career indicates the potential long-term benefits of the course.

The results also highlight some challenges and opportunities for educators. For instance, the student's preference for Japanese during group discussions indicates a need to create a more target-language-focused learning environment. Educators can consider providing additional English language support to encourage students to use their L2 more frequently. One potential approach to addressing this issue is to integrate Collaborative Online International Learning (COIL) into the course (Vahed & Rodriguez, 2020), thereby ensuring that students are exposed to the target language more extensively and are compelled to use it in meaningful communicative contexts.

Additionally, the importance of digital literacy skills in the current educational landscape cannot be overstated. The students' positive feedback on the digital aspects of the course highlights the potential for online education to enhance traditional academic skills and develop digital literacy skills that are becoming increasingly essential in today's society. As such, educators need to incorporate digital tools and skills into their curriculum to prepare students for the demands of the 21st-century workforce.

The immersive flipped course provided a gamified collaborative environment, which promoted student engagement and perceived learning outcomes. This observation is consistent with previous research findings indicating that the flipped classroom model can significantly increase students' engagement and motivation by allowing them to engage in active learning, collaborative work, and immediate feedback (Al-Samarraie et al., 2020; Nouri, 2016; Tseng et al., 2018; Wang, 2017).

Additionally, the immersive approach, which leverages the benefits of the Metaverse, has been found to promote effective collaboration. The results of the present study align with previous findings (Lee, 2021; Zou, 2020), as the students reported high levels of engagement and satisfaction with the immersive and gamified design of the course. This finding suggests that immersive flipped learning can be an effective instructional method to enhance student engagement and, potentially, learning outcomes.

This study adds to the growing body of evidence supporting the effectiveness of flipped classrooms, especially in online environments. It highlights the importance of technology in creating engaging and interactive learning experiences. Padlet, in particular, played a significant role in supporting student learning and interaction in a safe environment. By using Padlet, students could collaborate and share their ideas and insights, which helped them develop their critical thinking and communication skills.

Moreover, Padlet provided a platform for students to receive immediate feedback from their peers and the instructor, which improved their understanding of the course materials. Padlet also facilitated a safe and inclusive learning environment, as students could share their thoughts and opinions without fear of judgment or discrimination. As previous studies have shown (Ellis, 2015; Zou & Xie, 2018), Padlet proved to be an indispensable tool for stimulating discussion and ensuring all students understood the lesson before engaging in practice-oriented group work.

Reflection on learning has been a critical component of this innovative approach to teaching academic skills. The use of reflective reports and diaries is not new. It has been widely used in the literature to explore students' learning experiences and perceived learning outcomes—e.g., Blau et al. (2020). In this course, students were given a meta-assignment where they wrote an academic report on how they learned academic skills.

This not only allowed them to reflect on their past learning and consolidate their understanding but also helped the teacher understand their perspectives. There is no need to emphasize that it is imperative to assign tasks to students that are based on their personalized learning experiences, allowing them to express their opinions and arguments meaningfully. While the course has achieved notable successes, there is always room for improvement in any educational experience.

The author remains eager to explore additional ways to enhance student engagement and learning outcomes and encourage communication in the target language. Additionally, further research is needed to explore the scalability and sustainability of the immersive flipped learning approach, including its impact on other student populations and in different delivery formats.

Furthermore, with the emergence of generative AI technologies such as ChatGPT, the author is interested in exploring how the teaching of academic writing will evolve. She is committed to developing learning experiences that leverage these tools to scaffold student learning effectively.

The findings of the study have the following implications for educators.

1. Immersive flipped learning environments can be beneficial in designing practical online and blended courses that promote student engagement and collaboration.
2. In second-language contexts, additional English language support is necessary to encourage students to use their target language frequently.
3. Incorporating digital tools and skills into the curriculum is essential to preparing students for the demands of the 21st-century workforce.
4. Technology tools can be used to create engaging and interactive learning experiences, promote collaboration and communication, and facilitate a safe and inclusive learning environment.
5. Reflection on learning can be part of any course as it allows for identifying areas of improvement, reinforcing newly acquired knowledge, and developing a deeper understanding of the subject matter.

## CONCLUSION

This article presents an innovative approach to teaching academic skills to undergraduate Japanese students. Following an immersive flipped learning approach, students were provided with learning materials ahead of time and engaged in hands-on practice during synchronous online lessons hosted on a 2D Metaverse with a focus on academic skills such as reading, writing, and presentation. Group work was a pivotal component of the course, allowing the instructor to monitor students as they participated in group discussions on the course online platform. Course evaluations showed that most students had a positive experience and believed that their academic, collaboration, and digital literacy skills had improved.

One limitation of the study was the absence of a baseline group, which affected the generalizability of the results. Future research could enhance the quality of the current study by including a comparison group and improving the course through experimentation with diverse student demographics and exploring different delivery formats.

## ACKNOWLEDGEMENT

The Japan Society funds this research for the Promotion of Science Grants-in-Aid for Scientific Research (Project Number: 22K13756).

## APPENDIX

**ONLINE ENGAGEMENT SURVEY IN ENGLISH AND JAPANESE**

The survey items were divided into statements and how-often questions on a five-point Likert scale, as displayed in Table A1.

Table A1

*Five-point Likert Scale of Survey Items*

| Statements | | How Often Items | |
|---|---|---|---|
| **Statements (EN)** | **Statements (JP)** | **How Often Items** | **How Often Items (JP)** |
| 5 = strongly agree | 5 = とてもそう思う | 5 = Always | 5 = 常に |
| 4 = agree | 4 = そう思う | 4 = Usually | 4 = 大抵 |
| 3 = not sure | 3 = どちらともいえない | 3 = Sometimes | 3 = 時々 |
| 2 = disagree | 2 = あまりそう思わない | 2 = Rarely | 2 = ほとんどない |
| 1 = strongly disagree | 1 = 全くそう思わない | 1 = Never | 1 = 全くない |

*Course clarity*

Based on my experiences with and perceptions of this online course:
このオンライン授業のご自身の受講経験と認識について、以下の項目にお答えください。

**Item 1.** The organization of the course was clear.
授業の構成は明確だった。

**Item 2.** The instructions for the use of technology were clear.
オンラインツールの使い方の指示は分かりやすかった。

**Item 3.** The instructions for assignments were clear.
課題の指示は分かりやすかった。

**Item 4.** The course objectives were clear.
授業の目標は明確だった。

**Item 5.** The course content was explicit.
授業内容は明確だった。

*Student connectedness*

Based on my online class interactions with students in my class, I perceive:
このオンライン授業での他の受講生とのやりとりについて、以下の項目にお答えください。

**Item 6.** Students are respectful of one another.
受講生はお互いを尊重し合っていた。

**Item 7.** Students are cooperative with one another.
受講生は互いに協力し合っていた。

**Item 8.** Students are comfortable with one another.
受講生は仲が良かった。

**Item 9.** Students are supportive of one another.
受講生は互いに支え合っていた。

*Course structure*

Based on my experiences with and perceptions of this online course:
このオンライン授業のご自身の受講経験と認識について、以下の項目にお答えください。

**Item 10.** The design of this course encouraged student interaction.
このオンライン授業は、受講生同士の交流を促すような設計になっていた。

**Item 11.** This online course provided ample opportunities for communication among students.
このオンライン授業では、受講生同士のコミュニケーションの機会が十分にあった。

**Item 12.** The technology used in this course fostered collaboration among students.
このオンライン授業で活用されたオンラインツールは、受講生間のコラボレーションを促進した。

*Provision of Choice*

**Item 13.** How often did you get to choose your partners for online group work projects?
オンライングループワークで、グループのメンバーを自由に選ぶ機会はどの程度ありましたか？

**Item 14.** How often were students' ideas and suggestions used during online classroom discussions?

オンライン授業におけるディスカッションで、あなたのアイデアや意見はどの程度使われましたか？

**Item 15.** How often did students get to decide on group work details?

オンライングループワークで、グループワークの進め方に関してどれくらい自由に決めることができましたか？

**Item 16.** How often did students get to participate in making online classes' rules and policies?

このオンライン授業のルールや方針作りに受講者が参加する機会はどの程度ありましたか？

*Teaching relevance*

**Item 17.** How often did you discuss problems and issues that were meaningful to you?

あなたにとって有意義な問題や課題について、どれくらいの頻度で議論しましたか？

**Item 18.** How often did you learn things that were related to your life outside the online classroom?

このオンライン授業外で、自分の生活に関連したことを学ぶ機会はどの程度ありましたか？

**Item 19.** How often did you learn things that were useful for your future jobs?

将来の仕事に役立つ学びはどの程度ありましたか？

*Teacher emotional support*

Based on my online class interactions with the instructor, I perceived my instructor:

このオンライン授業における講師とのやりとりについて、以下の項目にお答えください。

**Item 20.** as understanding.

講師は理解を示してくれた。

**Item 21.** as respectful to me.

講師は私を尊重してくれた。

**Item 22.** as caring about me.

講師は気遣ってくれた。

*Teacher presence*

Based on my online class interactions with the instructor, I perceived my instructor:

このオンライン授業における講師とのやりとりについて、以下の項目にお答えください。

**Item 23.** as responsive.

講師の対応は良かった。

**Item 24.** as engaged in the course.

講師は授業に熱心に携わっていた。

**Item 25.** as approachable.

講師は親しみやすかった。